\documentclass[conference]{IEEEtran}

\usepackage{cite}
\usepackage{amsfonts, amssymb, amsthm, bm, bbm, graphicx, relsize, multirow, booktabs, tikz,subfigure,soul}
\usepackage{ucs}
\usepackage[cmex10]{amsmath,mathtools}
\usepackage[american]{babel}
\usepackage[T1]{fontenc}
\usepackage{tabulary}
\usepackage{algorithmic, algorithm}
\usepackage[multiple]{footmisc}

\newcommand{\beq}{\begin{equation}}
\newcommand{\eeq}{\end{equation}}
\newcommand{\bH}{\mathbf{H}}

\newcommand{\bD}{\mathbf{D}}

\newcommand{\bPhi}{\mathbf{\Phi}}

\newcommand{\bh}{\mathbf{h}}

\newcommand{\ds}{\displaystyle}
\newcommand{\norm}[1]{\left\lVert#1\right\rVert}    

\usepackage{graphicx}

\ifCLASSINFOpdf
\else
\fi
\usepackage[cmex10]{amsmath}
\usepackage{algorithmic}
\begin{document}
% No dashes to duplicated names in the references
\bstctlcite{IEEE_nodash:BSTcontrol}

\title{Resource Allocation in Wireless Networks Assisted by Reconfigurable Intelligent Surfaces}

\author{\IEEEauthorblockN{S. Buzzi, C. D'Andrea, A. Zappone}
\IEEEauthorblockA{University of Cassino and Southern Latium\\
Dept. of Electrical and Information Eng.\\
Cassino, Italy}
\and
\IEEEauthorblockN{M. Fresia}
\IEEEauthorblockA{Huawei Technol. Duesseldorf GmbH\\
Wireless Terminal Chipset Technology Lab\\
Munich, Germany}
\and
\IEEEauthorblockN{Y.-P. Zhang and S. Feng}
\IEEEauthorblockA{HiSilicon Technologies\\
Balong Solution Department\\
Bejing, China}}


\maketitle

\begin{abstract}
Reconfigurable Intelligent Surfaces (RISs) are recently attracting a wide interest due to their capability of tuning wireless propagation environments in order to increase the system performance of wireless networks. In this paper, a multiuser single-cell wireless network assisted by a RIS is studied. First of all, for the special case of a single-user system, three possible approaches are shown in order to optimize the Signal-to-Noise Ratio with respect to the beamformer used at the base station and to the RIS phase shifts. Then, for a multiuser system, assuming channel-matched beamforming, the geometric mean of the downlink Signal-to-Interference plus Noise Ratios across users is maximized with respect to the base stations transmit powers and   RIS phase shifts configurations. Numerical results show that the proposed procedure are effective and greatly improve the performance of the considered systems. 
\end{abstract}


% \fi
\IEEEpeerreviewmaketitle

\section{Introduction}
While massive MIMO has been a breakthrough technology that has significantly contributed to the evolution of wireless networks in the last decade,  new technologies and solutions have recently started to appear and gather attention as possible evolution of massive MIMO systems. These  include, among others, cell-free massive MIMO systems \cite{ngo2015cell,BuzziWCL2017}, the use of massive MIMO arrays for joint communication and sensing \cite{BuzziAsilomar2019}, large distributed antenna arrays \cite{amiri2018extremely}, and, also, reconfigurable intelligent surfaces (RISs) \cite{hu2017potential,hu2018beyond,di2019smart}. 
RISs are thin surfaces that can be used to coat buildings, ceilings, or other surfaces; they have electromagnetic properties that can be tuned electronically through a software controller, and their use permits modifying the propagation environment of wireless signals, so as to be able to concentrate information signals where needed and thus to improve the 
Signal-to-Interference plus Noise Ratio (SINR). 
Prototypes of reconfigurable metasurfaces are currently being developed in many parts of the world
\cite{VISOSURF_project,NTT_DoCoMo_LIS2019}.

Although being a quite recent topic, RISs have attracted an extraordinary interest in the wireless communications community. 
Paper \cite{wu2019beamforming} considers a multiple antenna transmitter and a single antenna receiver with a RIS, and formulates an optimization problem aiming at minimizing the transmit power under the assumption that  the phase shifts of the RIS tcan take values in a finite cardinality set. Paper \cite{huang2019reconfigurable} considers a multiple antenna transmitter and several single antenna receivers, and, under the assumption that (a) the direct path between the base station (BS) and the mobile stations (MSs) is blocked,  (b) perfect channel state information (CSI) is available, and (c) zero-forcing beamforming is used at the BS,  proposes a procedure aimed at maximizing the system energy efficiency through an alternating maximization approach.  Paper \cite{huang2018energy} considers a similar scenario in paper \cite{huang2019reconfigurable} but considers discrete phase shifts at the RIS. Paper \cite{basar2019transmission} considers a RIS assisted communication between a single antenna transmitter and receiver; under the assumption of perfect CSI,  the paper shows that RIS-based tranmission can effectively boost the received SNR and enable ultra-reliable communications at extremely low SNR values. Paper \cite{nadeem2019LIS} consider a multiuser scenario, with multiple antenna transmitter and single antenna receivers; for the case in which the direct BS-MS path is blocked, 
the optimal linear receiver that maximizes the downlink (DL) SINR is derived. 

Following on this track, this paper considers a single-cell wireless network assisted by a RIS. For a single user scenario, we maximize the signal-to-noise ratio (SNR) with respect to the choice of the beamformer at the BS and of the phase shifts at the RIS. Three different algorithms are proposed to this end, one based on a classical alternating-maximization approach, and two based on the maximization of a lower bound and of an upper bound of the SNR. For the latter two cases, the optimal solution is obtained in closed form. Next, for a multiuser scenario, we maximize the geometric mean of the DL SINRs (so as to take into account system fairness) with respect to the transmit power vector and to the RIS configuration. The alternating maximization approach is used in this case; in particular, the maximization with respect to the RIS phases is carried out through the gradient algorithm, while the maximization with respect to the transmit powers, after a proper reformulation of the problem in an equivalent convex form, is optimally solved at each iteration. One distinguishing feature of our study is that we consider the general cases that, for each MS, the direct BS-MS and the reflected BS-RIS-MS paths may be or may not be simultaneously active, whereas in many of the cited papers the assumption that the direct BS-MS path is blocked is crucial in order to solve the considered optimization problems.

This paper is organized as follows. Next section is devoted to the description of the system model. Section III and IV contain the description of the proposed optimization procedures, for the single-user and the multiuser case, respectively.
In Section V sample numerical results are presented, showing the effectiveness of the proposed procedures, while, finally concluding remarks are given in Section VI.

\section{System model}
\begin{figure}[!t]
\centering
\includegraphics[scale=0.6]{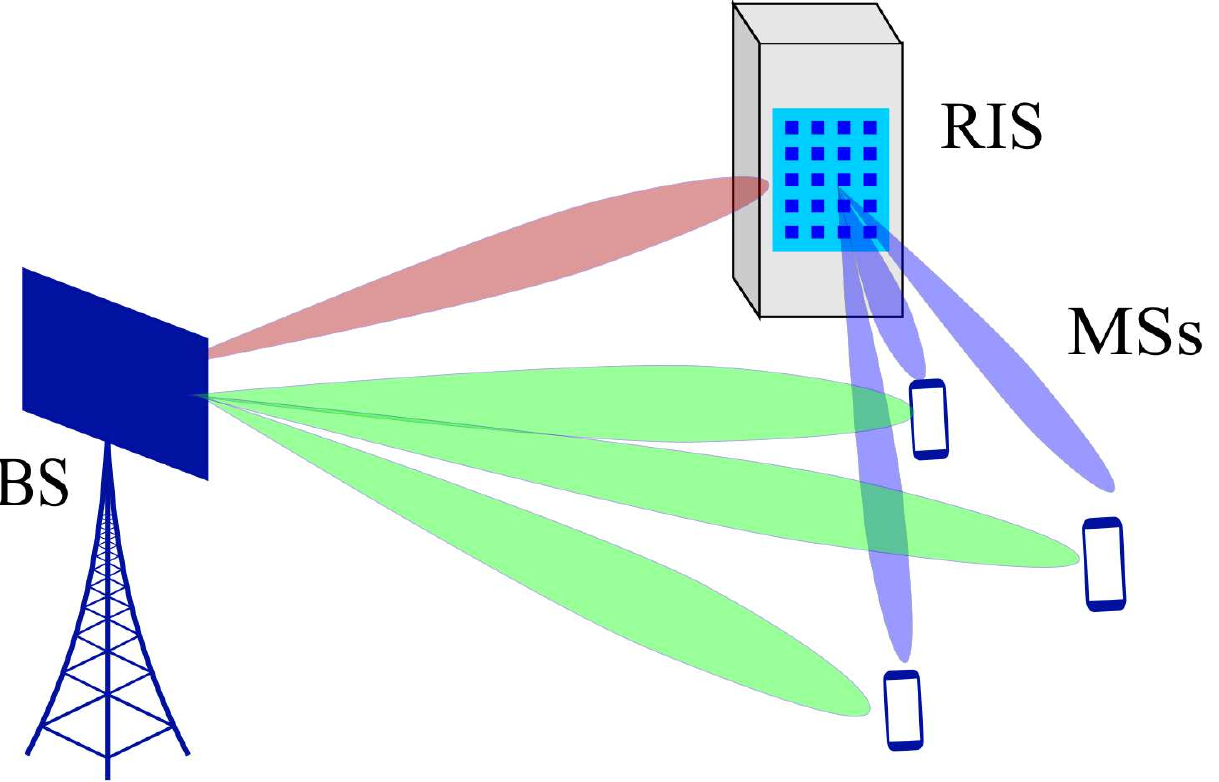}
\caption{RIS-assisted massive MIMO system. A base station equipped with a massive MIMO array communicates with a set of users in the same frequency band relaying both on a direct link and on a further link reflected by a planar array of reflecting devices with tunable phase reflection shift.}
\label{Fig:Scenario_3D}
\end{figure}

We consider a single-cell system with a BS equipped with $N_B$ antennas, a RIS with $N_R$ passive elements, and $K$ single antenna MSs,  as depicted in Fig. \ref{Fig:Scenario_3D}. 

We denote by $\bH$ the $(N_B \times N_R)$-dimensional matrix representing the wireless channel from the RIS to the BS, by $\bh_k$ the $N_R$-dimensional vector representing the channel from the $k$-th MS to the RIS, and by $\bh_{k,d}$ the $N_B$-dimensional vector representing the direct link channel from the $k$-th MS to the BS. Since some of the links between the MSs and the BS and the RIS may be blocked, we denote by $i_{k,d}$ and $i_{k,r}$ the binary
$0-1$  variables indicating the existence of a direct BS-MS and of a reflected BS-RIS-MS path for the $k$-th user, respectively. These variables $i_{k,d}$ and $i_{k,r}$, $\forall k$, are assumed to be deterministic, and depend on the geometry of the considered scenario.  
Accordingly,  ${\cal K}_d$ denotes the set of the MSs such that $i_{k,d}=1$ , while ${\cal K}_r$ denotes the set of the MSs such that $i_{k,r}=1$. 
The RIS behaviour is modeled through a set of $N_R$ complex coefficients, representing the loss and the phase shift imposed on reflected waves by the RIS elements. These coefficients are arranged into the matrix 
$
\bPhi= \textrm{diag}\left(\rho e^{j\phi_1}, \ldots, \rho e^{j\phi_{N_R}}\right)
$.
The positive real-valued coefficient $\rho$ accounts for possible reflection losses and is constant across the RIS elements, while the phase offsets $\phi_1, \ldots, \phi_{N_R}$ can be software-controlled by the network.

The composite channel from the $k$-th MS to the BS, when the RIS configuration is given by the matrix $\bPhi$ is written as
\begin{equation}
\overline{\bh}_{k, \bPhi}=\sqrt{\beta_k}\bH \bPhi \bh_k i_{k,r} +
\sqrt{\beta_{k,d}}\bh_{k,d}i_{k,d},
\label{comp_channel}
\end{equation}
where $\beta_k$ and $\beta_{k,d}$ denote the $k$-th MS power attenuation coefficients 
for the reflected MS-RIS-BS path and for the direct MS-BS path, respectively. The Time Division Duplex (TDD) protocol is used so that the DL channel is the same as the uplink. 
%\vspace{-0.1cm}

Focusing on DL data transmission, let us denote by  $\eta_k^{\rm DL}$ the DL transmit power
reserved for the $k$-th MS, by $q_k^{\rm DL}(n)$  the information symbol transmitted by the BS and intended for the $k$-th MS in the $n$-th (discrete) symbol interval. Additionally, let  $\mathbf{w}_k$ be the $N_B$-dimensional beamforming vector for the $k$-th MS. Upon transmission of the data multiplex from the BS, the baseband equivalent of the discrete-time observable at the $k$-th MS antenna, coinciding with  the soft estimate of the information symbol 
$q_k^{\rm DL}(n)$ can be written as 
\begin{equation}
\begin{array}{lll}
\widehat{q}_k^{\rm DL}(n)=&\sqrt{\eta_k^{\rm DL}}\overline{\bh}_{k, \bPhi}^H \mathbf{w}_k  q_k^{\rm DL}(n) \\ &+ \ds \sum_{\substack{j=1 \\ j \neq k}}^K {\sqrt{\eta_j^{\rm DL}} \overline{\bh}_{k, \bPhi}^H \mathbf{w}_j q_j^{\rm DL}(n)} + z_k(n) \, .
\end{array}
\label{DL_signal_k}
\end{equation}
where $z_k(n) \sim \mathcal{CN}(0,\sigma^2_z)$ is the AWGN contribution at the $k$-th MS receiver in the $n$-th symbol interval.
Based on \eqref{DL_signal_k}, we are now ready to detail the resource allocation procedures for the single-user and the multi-user scenario.

\section{Single user closed-form joint active and passive beamforming design} \label{Single_user_Resource}
This section focuses on the special case of a single-user system, which may be representative of a 
network with an orthogonal multiple access scheme is used and with negligible co-channel interference. 
We tackle the maximization of the system SNR with respect to the base station beamforming vector $\mathbf{w}$ (active beamforming) and of the RIS phase shifts (passive beamforming). In the following, let us denote by $P_T$ the BS transmit power during the data transmission phase, by $q^{\rm DL}(n)$ the information symbol intended for the MS in the $n$-th (discrete) symbol interval, which can be expressed as 
\begin{equation}
\begin{array}{lll}
\widehat{q}^{\rm DL}(n)=\sqrt{P_T} \left( \bH \bPhi \bh +\bh_{d} \right)^H \mathbf{w}  q^{\rm DL}(n) + z(n) \, ,
\end{array}
\label{DL_signal_k_SU}
\end{equation}
with $z(n) \sim \mathcal{CN}(0,\sigma^2_z)$ denoting thermal noise.

Our first step, is to rewrite \eqref{DL_signal_k_SU} in a more convenient form, by exploiting the following identity
\beq
\bH \bPhi \bh_k= \bD_k  \bm{\phi}\; ,
\label{eq:identity}
\eeq
where $\bD_k(i,j) \triangleq \bH(i,j) \bh_k(j)$, for all $i=1, \ldots, N_B$, $j=1, \ldots, N_L$ and $\bm{\phi}=\textrm{diag}(\bPhi)$ is the column $N_L$-dimensional vector containing the diagonal entries of $\bPhi$, which enables us to rewrite \eqref{DL_signal_k_SU} as
\begin{equation}
\begin{array}{lll}
\widehat{q}^{\rm DL}(n)=\sqrt{P_T} \left( \bD  \bm{\phi} +\bh_{d} \right)^H \mathbf{w}  q^{\rm DL}(n) + z(n) \, ,
\end{array}
\label{DL_signal_k2_SU}
\end{equation}
Based on \eqref{DL_signal_k_SU}, the system SNR can be defined as
\begin{equation}
\text{SNR}= \frac{P_T}{\sigma_z^2} \left| \mathbf{w}^H \left( \bD  \bm{\phi} +\bh_{d} \right) \right|^2
\end{equation}
In practice, the BS will have access to estimates of $\bD$ and $\bh_{d}$, which will be denoted in the following by $\widehat{\bD}$ and $\widehat{\bh}_{d}$, respectively, which implies that the function that can be optimized at the transmit side is 
\begin{equation}
\widehat{\text{SNR}}= \frac{P_T}{\sigma_z^2} \left| \mathbf{w}^H \left( \widehat{\bD}  \bm{\phi} +\widehat{\bh}_{d} \right) \right|^2
\end{equation}
Then, the problem to solve is stated as 
\begin{subequations}\label{Prob:MaxSNR}
 \begin{align}
&\ds\max_{\mathbf{w}, \bm{\phi}}\; \; \; \;  \left| \mathbf{w}^H \left( \widehat{\bD}  \bm{\phi} +\widehat{\bh}_{d} \right) \right|^2  \label{Prob:Max_SNR}\\
&\;\textrm{s.t.}\; \; [\bm{\phi}]_i= \rho e^{j \phi_{i}},  \\
&\;\; \; \;\; \; \; \phi_{i} \, \in \, [-\pi, \pi], \, \forall \; i=1, \ldots, N_R \\
&\;\; \;\;\; \; \; \|\mathbf{w}\|^2=1
\end{align}
\end{subequations}
Problem of the form of \eqref{Prob:MaxSNR} are usually tackled by alternating optimization methods which iterate between the optimization of the base station beamforming vector $\mathbf{w}$ and of the RIS phase shifts $\bm{\phi}$. This approach could be used also for the case at hand, but it has the drawback of requiring a numerical iterative algorithm. Instead, in the following, we propose two optimization methods that optimize an upper-bound and a lower-bound of the objective of \eqref{Prob:MaxSNR}, and which have the advantage of leading to closed-form expressions of $\mathbf{w}$ and $\bm{\phi}$. 

\subsection{Upper-bound maximization} \label{UB_max_Section}
Assume, without loss of generality, that $N_B<N_R$, and consider the singular value decomposition of $\widehat{\bD}$, i.e.,
\begin{equation}
\widehat{\bD}=\sum_{i=1}^{N_B} \lambda_i \mathbf{u}_i \mathbf{v}_i^H\;.
\end{equation}
Next, let us express $\widehat{\bh}_{d}$ in terms of its projection on the orthonormal basis vectors $\mathbf{u}_1, \ldots, \mathbf{u}_{N_B}$, i.e., 
\begin{equation}
\widehat{\bh}_{d}=\sum_{i=1}^{N_B} \alpha_i \mathbf{u}_i, 
\end{equation}
where $\alpha_i=\mathbf{u}_i^H\widehat{\bh}_{d}$. 
 
At this point, we observe that an upper bound of the objective of Problem \eqref{Prob:MaxSNR} can be written as
\begin{equation}
\begin{array}{lllll}
\left| \mathbf{w}^H \left( \widehat{\bD}  \bm{\phi} +\widehat{\bh}_{d} \right) \right|^2 &= \left| \mathbf{w}^H \left[ \ds \sum_{i=1}^{N_B} \mathbf{u}_i \left( \lambda_i \mathbf{v}_i^H \bm{\phi} +\alpha_i\right) \right] \right|^2 \\ & \leq N_B \ds  \sum_{i=1}^{N_B} \left|\mathbf{w}^H\mathbf{u}_i\right|^2 \left| \lambda_i \mathbf{v}_i^H \bm{\phi} +\alpha_i \right|^2
\end{array}
\label{SNR}
\end{equation} 
The upper-bound in \eqref{SNR} can be jointly maximized with respect to both $\bm{\phi}$ and $\mathbf{w}$. To see this, let us first consider, for all $i=1,\ldots,N_{B}$, the following optimization problem
\begin{equation}
\ds\max_{\bm{\phi}} \left| \lambda_i \mathbf{v}_i^H \bm{\phi} +\alpha_i \right|^2 = \ds\max_{\bm{\phi} } \left| \lambda_i \mathbf{v}_i^H \bm{\phi} e^{-j \angle{\alpha_i}} +|\alpha_i| \right|^2\;,
\label{max_phi_i_UB}
\end{equation}
whose optimal solution $\bm{\phi}_i^{\rm opt}$ is found by noticing that the phase of the $n$-th entry of $\bm{\phi}_i^{\rm opt}$, say $\phi_{n,i}^{\rm opt}$, is given by 
\begin{equation}
\phi_{n,i}^{\rm opt}= -\angle{[\mathbf{v}_i^*]_n} + \angle{\alpha_i}\;.
\end{equation}
Next, let us define the index $i^{+}=\text{argmax}_{i}\left| \lambda_i \mathbf{v}_i^H \bm{\phi}_i^{\rm opt} +\alpha_i \right|^2$ and $c_{i^{+}}=\left| \lambda_{i^{+}} \mathbf{v}_{i^{+}}^H \bm{\phi}_{i^{+}}^{\rm opt} +\alpha_{i^{+}}\right|^2$. Thus, it follows that 
\begin{align}
\ds  \sum_{i=1}^{N_B} \left|\mathbf{w}^H\mathbf{u}_i\right|^2 \left| \lambda_i \mathbf{v}_i^H \bm{\phi} +\alpha_i \right|^2&\leq\notag\\ 
c_{i^{+}}\sum_{i=1}^{N_{B}}|\mathbf{w}^{H}\mathbf{u}_{i}|^{2}&\leq c_{i^{+}}\;,\label{Eq:UpperBound}
\end{align}
where we have also exploited that fact that both $\mathbf{w}^{H}$ and $\mathbf{u}_{i}$ have unit-norm. Finally, we observe that all inequalities in \eqref{Eq:UpperBound} turn to equalities by choosing 
$
\bm{\phi}^{\rm opt}=\bm{\phi}_{i^+}^{\rm opt}$ and $\mathbf{w}^{\rm opt}=\mathbf{u}_{i^+}$,
which therefore are the maximizers of the right-hand-side of \eqref{SNR}. 
 
\subsection{Lower-bound Maximization} \label{LB_Max_Section}
Define $\mathbf{g}_{w}^H=\mathbf{w}^H \widehat{\bD}$, and $t_{w}=\mathbf{w}^H\widehat{\bh}_{d}$. Then, it holds:
%the objective of Problem \eqref{Prob:MaxSNR} can be lower-bounded as
\begin{align}
&\ds\max_{\mathbf{w},\bm{\phi}}\left|\mathbf{w}^{H}(\widehat{\bD}\bm{\phi}+\widehat{\bh}_{d})\right|^{2}= \ds \max_{\mathbf{w}}\left(\max_{\bm{\phi}}\left|\mathbf{g}_{w}^{H}\bm{\phi}+t_{w})\right|^{2}\right)\stackrel{(a)}{=}\notag\\
&\ds\rho^2 \max_{\mathbf{w}}\left|\sum_{i=1}^{N_{R}}|\mathbf{g}_{w}(i)|+|t_{w}|\right|^{2}\stackrel{(b)}\geq
\ds \rho^2\max_{\mathbf{w}}\left|\mathbf{w}^{H}\left(\sum_{i=1}^{N_{R}}\widehat{\mathbf{d}}_{i}+\widehat{\bh}_{d}\right)\right|^{2}
\label{SNR_LB}
\end{align} 
where the equality $(a)$ holds since, for any given $\mathbf{w}$, the optimal $\bm{\phi}$ is the one that aligns the phases of $t_{w}$ and of the components of $\mathbf{g}_{w}$, denoted by $\mathbf{g}_{w}(i)$ with $i=1,\ldots,N_{R}$, while inequality (b) holds by removing the inner absolute values and since $\mathbf{g}_{w}(i)=\mathbf{w}^{H}\widehat{\mathbf{d}}_i$, with $\widehat{\mathbf{d}}_i$ the  $i$-th column of $\widehat{\bD}$.

From \eqref{SNR_LB}, we see that the optimal $\mathbf{w}$ has the form:
\begin{equation}
\mathbf{w}^{\rm opt}=\frac{\ds \sum_{i=1}^{N_B} \widehat{\mathbf{d}}_i + \widehat{\bh}_{d}}{\norm{\ds \sum_{i=1}^{N_B} \widehat{\mathbf{d}}_i + \widehat{\bh}_{d}}}\;,
\end{equation}
from which we can obtain the optimal phases of the RIS as  
%\begin{equation}
%\ds\max_{\bm{\phi}} \left| \mathbf{g}^H \bm{\phi} +t \right|^2 = \ds\max_{\bm{\phi} } \left| \mathbf{g}^H \bm{\phi}e^{-j \angle{t}} +|t| \right|^2,
%\label{max_phi_i_LB}
%\end{equation}
%i.e., denoting as $\phi_{n}^{\rm opt}$ the phase of the $n$-th entry of the optimum $\bm{\phi}$ 
\begin{equation}
\phi_{i}^{\rm opt}= -\angle{\mathbf{g}_{w}^{*}(i)} + \angle{t}\;,\;\forall\;n=1,\ldots,N_{R}
\end{equation}

\section{Multiuser joint passive beamforming design and power allocation} \label{Joint_Resource}
In the general multi-user scenario, the problem to solve is stated as 
\begin{subequations}\label{Prob:MaxSNR_MU}
 \begin{align}
& \ds\max_{\bm{\eta}^{\rm DL}, \bm{\phi}} \; \prod_{k=1}^K \frac{\eta_k^{\rm DL}\left| \left( \widehat{\bD}_k  \bm{\phi}  +\widehat{\bh}_{k,d}\right)^H\mathbf{w}_k\right|^2}{\ds \sum_{\substack{\ell=1 \\ \ell \neq k}}^K {\eta_{\ell}^{\rm DL}\left| \left( \widehat{\bD}_k  \bm{\phi}  +\widehat{\bh}_{k,d}\right)^H\mathbf{w}_{\ell}\right|^2}  +   \sigma^2_z} \, , \\
&\;\textrm{s.t.}\; [\bm{\phi}]_i= \rho e^{j \phi_{i}},  \\
&\; \; \; \; \; \phi_{i} \, \in \, [-\pi, \pi], \, \forall \; i=1, \ldots, N_R \\
&\; \; \; \; \; \sum_{\ell=1}^K {\eta_{\ell}^{\rm DL}} \leq P_{\rm max}^{\rm BS}, \\
&\; \; \; \; \; {\eta_{\ell}^{\rm DL}} \geq 0 \; \forall \ell=1,\ldots,K ,  
\end{align}
\end{subequations}

where $\bm{\eta}^{\rm DL}= \left[ \eta_1^{\rm DL}, \ldots, \eta_K^{\rm DL}\right]^T$ and we are assuming again that the BS treats the channel estimates as the true channels. Substituting $\bm{\phi}= \rho e^{j \bm{\widetilde{\phi}}}$, with $\bm{\widetilde{\phi}}= \left[ \phi_1, \ldots \ \phi_{N_R}\right]^T$, and assuming channel-matched beamforming (CM-BF), i.e., the $k$-th MS pre-coding beamforming vectors are chosen as
\begin{equation}
\mathbf{w}_k= \displaystyle \frac{\rho \widehat{\bD}_{k} e^{j\bm{\widetilde{\phi}}}+\widehat{\bh}_{k,d}}{\norm{\rho \widehat{\bD}_{k} e^{j\bm{\widetilde{\phi}}}  +\widehat{\bh}_{k,d}}}  \; ,
\label{CM_BF}
\end{equation} 
Problem \eqref{Prob:MaxSNR_MU} can be rewritten as 
\begin{subequations}\label{Prob:MaxSNR_MU2}
 \begin{align}
&\ds\max_{\bm{\eta}^{\rm DL}, \bm{\widetilde{\phi}}} \; \prod_{k=1}^K \frac{\eta_k^{\rm DL}\norm{\rho \widehat{\bD}_k e^{j \bm{\widetilde{\phi}}}  +\widehat{\bh}_{k,d}}^2}{\ds \sum_{\substack{\ell=1 \\ \ell \neq k}}^K {\eta_{\ell}^{\rm DL}\frac{\left|\left( \rho \widehat{\bD}_k e^{j\bm{\widetilde{\phi}}}\!\!  +\!\!\widehat{\bh}_{k,d}\right)^{\!\!H}\!\!\!\!\left( \rho \widehat{\bD}_{\ell} e^{j\bm{\widetilde{\phi}}} \!\! +\!\!\widehat{\bh}_{\ell,d} \right)\right|^2}{\norm{\rho \widehat{\bD}_{\ell} e^{j\bm{\widetilde{\phi}}}  +\widehat{\bh}_{\ell,d}}^2}} \!\! + \!\!  \sigma^2_z} \, , \\
&\; \;\; \;\text{s. to} \;  \left[\bm{\widetilde{\phi}}\right]_{i} \, \in \, [-\pi, \pi], \, \forall \; i=1, \ldots, N_R , \\
&\;\;\;\; \; \; \; \;  \sum_{\ell=1}^K {\eta_{\ell}^{\rm DL}} \leq P_{\rm max}^{\rm BS}, \\
&\;\;\;\; \; \; \; \;  {\eta_{\ell}^{\rm DL}} \geq 0 \; \forall \ell=1,\ldots,K .  
\end{align}
\end{subequations}
Solving \eqref{Prob:MaxSNR_MU2} optimally appears challenging, due to the presence of multi-user interference. This motivates us to resort to alternating optimization to find a candidate solution of  \eqref{Prob:MaxSNR_MU2} \cite[Section 2.7]{BertsekasNonLinear}, i.e, we solve alternatively the problem with respect to $\bm{\widetilde{\phi}}$ and then the problem with respect to $\bm{\eta}^{\rm DL}$. At each step, the objective of \eqref{Prob:MaxSNR_MU2} does not decrease, and so the algorithm converges in the value of the objective function.

\subsection{Solution of the problem with respect to $\bm{\widetilde{\phi}}$} \label{Phi_opt_Section}
Determining the optimal solution of Problem \eqref{Prob:MaxSNR_MU2} appears challenging even with respect to only the RIS phases, due to the fact multiple users are present, that are served by the same RIS matrix $\bm{\phi}$. This prevents from obtaining significant insight on the optimal $\bm{\Phi}$. On the other hand, it was shown in \cite{huang2019reconfigurable} that good results are obtained when employing a gradient-based search to optimize the phase shift matrix in RIS-based networks. Here we take a similar approach, applying the gradient algorithm in order to find a candidate solution for Problem \eqref{Prob:MaxSNR_MU2}. Before applying the gradient algorithm, we equivalently reformulate the problem by taking the logarithm of the objective, which leads us to Problem \eqref{Prob:MaxSNR_MU2_phi}, shown at the top of next page\footnote{Without loss of generality, we have neglected the constraint $\left[\bm{\widetilde{\phi}}\right]_{i} \, \in \, [-\pi, \pi], \, \forall \; i=1, \ldots, N_R$, since the objective is periodic with respect to each phase, with period $2\pi$, and thus any phase can be restricted to this fundamental period after the optimization routine has converged.}. This entails no loss of optimality, due to the fact that the logarithm is an increasing function. At this point, denoting by $G(\bm{\widetilde{\phi}})$ the objective of \eqref{Prob:MaxSNR_MU2_phi}, and by $\widetilde{\phi}_{n}$ the $n$-th component of $\bm{\widetilde{\phi}}$, for any  $n=1,\ldots,N_{R}$ it holds
\begin{align}
&\frac{\partial G}{\partial \widetilde{\phi}_{n}}=\log_{2}(e)\sum_{k=1}^{K}\sum_{\ell\neq k}\left(\frac{|F_{k,\ell}|^{2}}{F_{\ell,\ell}F_{k,k}}+\frac{\sigma_{z}^{2}}{F_{k,k}}\right)\times\\
&\!\frac{\ds\frac{\partial F_{k,k}}{\partial \widetilde{\phi}_{n}}\!\!\left(\!\ds\sum_{\ell\neq k}\!\frac{|F_{k,\ell}|^{2}}{F_{\ell,\ell}}\!+\!\sigma^{2}_{z}\!\right)\!\!-\!F_{k,k}\!\!\left(\!\ds\sum_{\ell\neq k}\!\frac{\partial |F_{k,\ell}|^{2}}{\partial \widetilde{\phi}_{n}}\!\frac{1}{F_{\ell,\ell}}\!-\!\frac{\partial F_{\ell,\ell}}{\partial\phi_{n}}\frac{|F_{k,\ell}|^{2}}{F_{\ell,\ell}^{2}}\!\!\right)}{\left(\ds\sum_{\ell\neq k}\frac{|F_{k,\ell}|^{2}}{F_{\ell,\ell}}+\sigma^{2}_{z}\right)^{2}}\notag,
\end{align}
wherein
\begin{align}
F_{k,\ell}&=\eta_{\ell}^{DL}\left( \rho \widehat{\bD}_k e^{j\bm{\widetilde{\phi}}} +\widehat{\bh}_{k,d}\right)^{H}\left( \rho \widehat{\bD}_{\ell} e^{j\bm{\widetilde{\phi}}}  +\widehat{\bh}_{\ell,d} \right)\\
\frac{\partial |F_{k,\ell}|^{2}}{\partial \widetilde{\phi}_{n}}&=2\Re\left\{\frac{\partial F_{k,\ell}}{\partial 
\widetilde{\phi}_{n}}F_{k.\ell}^{*}\right\}\\
\frac{\partial F_{k,\ell}}{\partial \phi_{n}}&=\eta_{\ell}^{DL}\rho j\Bigg[\rho \sum_{m\neq n}\left[\mathbf{\widehat{D}}_{k}^{H}\mathbf{\widehat{D}}_{\ell}\right]_{m,n}e^{j(\widetilde{\phi}_{n}-
\widetilde{\phi}_{m})}\\
&-\rho \sum_{i\neq n}\left[\mathbf{\widehat{D}}_{k}^{H}\mathbf{\widehat{D}}_{\ell}\right]_{n,i}e^{-j(\widetilde{\phi}_{n}-\widetilde{\phi}_{i})}\notag\\
&+e^{j\widetilde{\phi}_{n}}\left[\mathbf{\widehat{D}}_{\ell}^{T}\mathbf{\widehat{h}}_{k,d}^{*}\right]_n-e^{-j\widetilde{\phi}_{n}}\left[\mathbf{\widehat{D}}_{k}^{H}\mathbf{\widehat{h}}_{\ell,d}\right]_n \Bigg]\notag
\end{align}
Equipped with the expression of the derivatives of the objective of \eqref{Prob:MaxSNR_MU2_phi} with respect to all optimization variables, the well-known gradient algorithm can be implemented, also resorting to off-the-shelf software packages. 
\begin{figure*}
\begin{subequations}\label{Prob:MaxSNR_MU2_phi}
 \begin{align}
&\ds\max_{\bm{\widetilde{\phi}}} \sum_{k=1}^K \log_2 \left(\frac{\eta_k^{\rm DL}\norm{\rho \widehat{\bD}_k e^{j \bm{\widetilde{\phi}}}  +\widehat{\bh}_{k,d}}^2}{\ds \sum_{\substack{\ell=1 \\ \ell \neq k}}^K {\eta_{\ell}^{\rm DL}\frac{\left|\left( \rho \widehat{\bD}_k e^{j\bm{\widetilde{\phi}}} +\widehat{\bh}_{k,d}\right)^{H}\left( \rho \widehat{\bD}_{\ell} e^{j\bm{\widetilde{\phi}}}  +\widehat{\bh}_{\ell,d} \right)\right|^2}{\norm{\rho \widehat{\bD}_{\ell} e^{j\bm{\widetilde{\phi}}}  +\widehat{\bh}_{\ell,d}}^2}}  +  \sigma^2_z}\right) \, , 
%&\; \;\; \;\text{s. to} \;  \left[\bm{\widetilde{\phi}}\right]_{i} \, \in \, [-\pi, \pi], \, \forall \; i=1, \ldots, N_R. 
\end{align}
\end{subequations}
\end{figure*}

\subsection{Solution of the problem with respect to $\bm{\eta}^{\rm DL}$} \label{Eta_Opt_Section}
After the solution of Problem \eqref{Prob:MaxSNR_MU2_phi}, we solve the problem with respect to the variables $\bm{\eta}^{\rm DL}$ for fixed $\bm{\widetilde{\phi}}$. To begin with, we take the logarithm of the objective function of \eqref{Prob:MaxSNR_MU2_phi}, which causes no optimality loss, since the logarithm is an increasing function. Thus, the optimization problem with respect to the transmit powers can be equivalently reformulated as 

\begin{subequations}\label{Prob:MaxSNR_MU2_eta}
 \begin{align}
&\ds\max_{\bm{\eta}^{\rm DL}} \sum_{k=1}^K \ds  \log_2 \left(\frac{\eta_k^{\rm DL} a_{k,k}}{\ds \sum_{\substack{\ell=1 \\ \ell \neq k}}^K {\eta_{\ell}^{\rm DL} a_{k,\ell}}  +  \sigma^2_z}\right) \, , \\
&\; \;\; \;\text{s. to} \;  \sum_{\ell=1}^K {\eta_{\ell}^{\rm DL}} \leq P_{\rm max}^{\rm BS}, \\
&\;\;\;\; \; \; \; \;  {\eta_{\ell}^{\rm DL}} \geq 0 \; \forall \ell=1,\ldots,K ,  
\end{align}
\end{subequations} 
with
\begin{equation}
a_{k,\ell}= \ds \frac{\left|\left( \rho \widehat{\bD}_k e^{j\bm{\widetilde{\phi}}} +\widehat{\bh}_{k,d}\right)^{H}\left( \rho \widehat{\bD}_{\ell} e^{j\bm{\widetilde{\phi}}}  +\widehat{\bh}_{\ell,d} \right)\right|^2}{\norm{\rho \widehat{\bD}_{\ell} e^{j\bm{\widetilde{\phi}}}  +\widehat{\bh}_{\ell,d}}^2} \, .
\end{equation}
Next, we apply the change of variable 
$
\eta_{\ell}^{\rm DL}=2^{\gamma_{\ell}}, 
$
also defining $\bm{\gamma}=\left[ \gamma_1, \ldots, \gamma_K\right]^T$, which allows reformulating Problem \eqref{Prob:MaxSNR_MU2_eta} in the equivalent form
\begin{equation}\label{Prob:MaxSNR_MU2_eta2}
\ds\max_{\bm{\gamma}} \sum_{k=1}^K \ds \gamma_k + \log_2(a_{k,k}) -\log_2 \left(\ds \sum_{\substack{\ell=1 \\ \ell \neq k}}^K {2^{\gamma_{\ell}} a_{k,\ell}}  +  \sigma^2_z\right) \, , 
\end{equation} 
subject to $\;  \sum_{\ell=1}^K {2^{\gamma_{\ell}}} \leq P_{\rm max}^{\rm BS}$.
The advantage of our approach is that the non-convex problem \eqref{Prob:MaxSNR_MU2_phi} has been equivalently restated as Problem \eqref{Prob:MaxSNR_MU2_eta2}, which can be seen to have a concave objective function and a convex constraint function. Thus, \eqref{Prob:MaxSNR_MU2_eta2} is convex and can be optimally solved in polynomial time by standard convex optimization methods.
\begin{table}
\centering
\caption{Simulation parameters}
\label{table:parameters}
\def\arraystretch{1.2}
\begin{tabulary}{\columnwidth}{ |p{3cm}|p{4.5cm}| }
\hline
  MSs distribution 				& Horizontal: uniform, vertical: 1.5~m\\ \hline
  BS height				& 25~m\\ \hline
  RIS height				& 40~m\\ \hline
%  	Path-loss models & Eq. \eqref{Beta_k} \\ \hline
	Carrier freq., bandwidth		&  $f_0=3$ GHz, $B = 20$ MHz \\ \hline
	BS antenna array			& 16-element with $\lambda/2$ spacing\\ \hline
	RIS antenna array			& 32-element with $\lambda/2$ spacing\\ \hline
	MS antennas 		& Omnidirectional with 0~dBi gain\\ \hline% and vertical polarization  \\ \hline
	Thermal noise 				& -174 dBm/Hz spectral density \\ \hline
	Noise figure 			& 9 dB at BS/MS\\ \hline
\end{tabulary}
\end{table}

\begin{figure}[!t]
\centering
\includegraphics[scale=1.8]{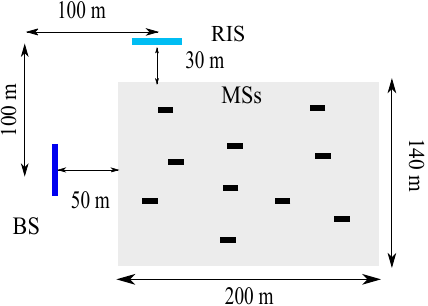}
\caption{The considered scenario.}
\label{Fig:Scenario2}
\end{figure}

\section{Numerical Results}

For thenumerical results, we refer to the scenario depicted in Fig. \ref{Fig:Scenario2}, which shows the position of the BS planar antenna array, of the RIS, and of the grey area where the  MSs are uniformly distributed. We consider a BS with $N_B=16$ antennas placed at a height of 25 m., and a RIS with  $N_R=32$ reflecting elements placed at a height of 40 m.; the number of MSs simultaneously served on the same frequency is $K=10$ in the multiuser scenario. The other simulation parameters are summarized in Table \ref{table:parameters}.

The channel attenuation coefficients $\beta_k$ and $\beta_{k,d}$ that model the attenuation on the reflected path $k$-th MS-RIS-BS and on the direct $k$-th MS-BS path, respectively, are modeled according to \cite{huang2018energy,huang2019reconfigurable}. In particular,
\begin{equation}
\beta_k=\frac{10^{-3.53}}{\left(d_{{\rm BS,RIS}}+d_{{\rm RIS},k}\right)^{3.76}} \, , \; \; \text{and} \; \; \beta_{k,d}=\frac{10^{-3.53}}{d_{{\rm BS},k}^{3.76}},
\label{Beta_k}
\end{equation}
where $d_{{\rm BS,RIS}}$ and $d_{{\rm RIS},k}$ are the distance between the BS and the RIS and the distance between the $k$-th MS and the RIS in meters, respectively, while $d_{{\rm BS},k}$ is the distance between the BS and the $k$-th MS. 

\begin{figure}[!t]
\centering
\includegraphics[scale=0.45]{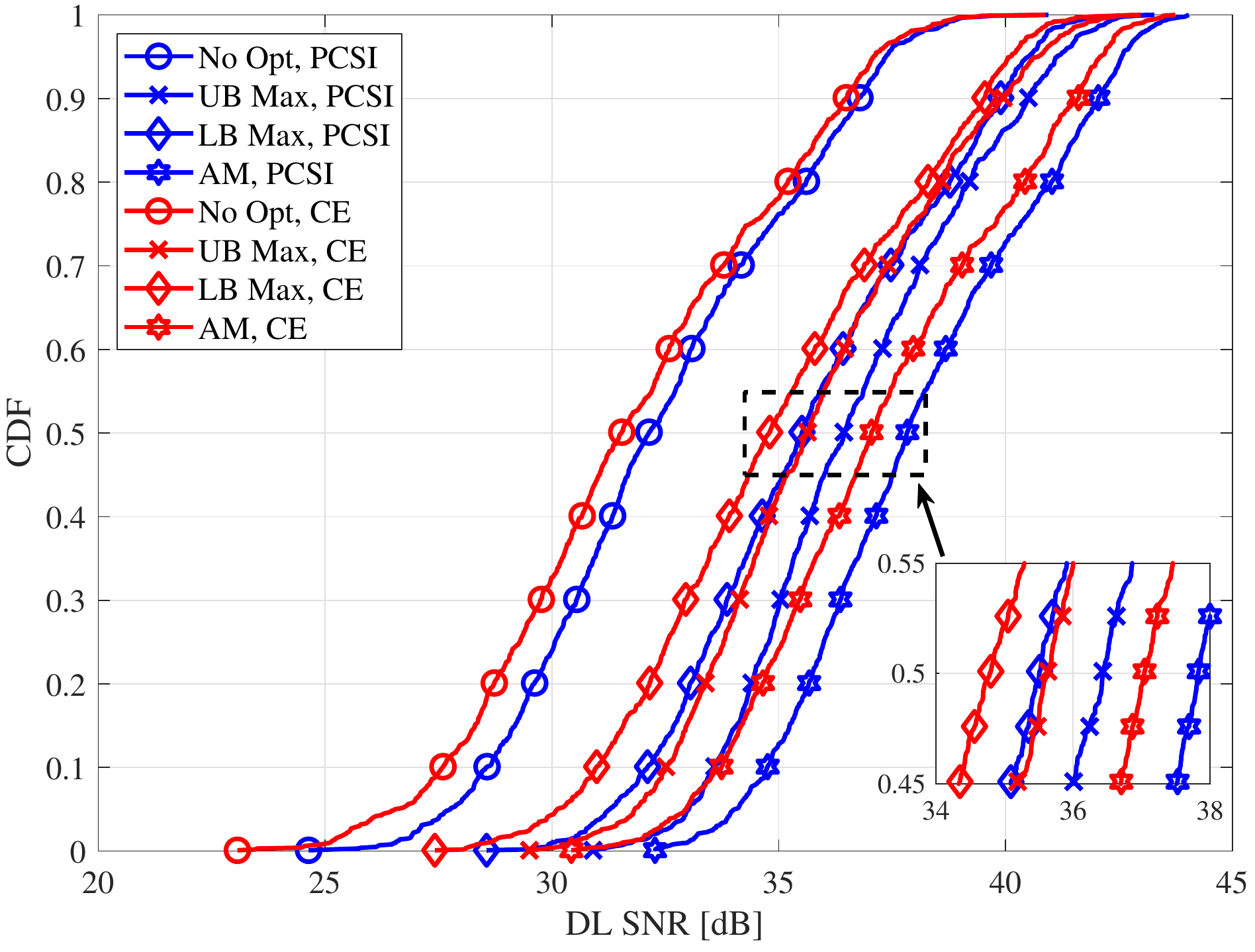}
\caption{CDFs of the SNR in the single user scenario.}
\label{Fig:SNR_SingleUser_Opt}
\end{figure}

We assume that the maximum power transmitted by the BS is $P_{\rm max}^{\rm BS}=10$ W. In the following figures, we report the results of the resource allocation strategies proposed in the paper in the case of PCSI and in the case in which channel estimation (CE) is performed at the BS. For the sake of brevity, details on the channel estimation procedure are omitted and will be possibly reported elsewhere. In Fig. \ref{Fig:SNR_SingleUser_Opt}, we report the CDFs of the SNR in the single user case when we use the optimization strategies reported in Section \ref{Single_user_Resource}. In particular, we compare the results of the upper-bound maximization (UB Max) in Section \ref{UB_max_Section} and of lower-bound maximization (LB Max) in Section \ref{LB_Max_Section} with the alternating maximization (AM) approach and with the case in which a random configuration is used for the RIS (No Opt.). We can see that the proposed strategies are effective and give performance very close to the AM with a significant lower complexity, given the fact that we have been able to express the solution in closed form.

\begin{figure}[!t]
\centering
\includegraphics[scale=0.45]{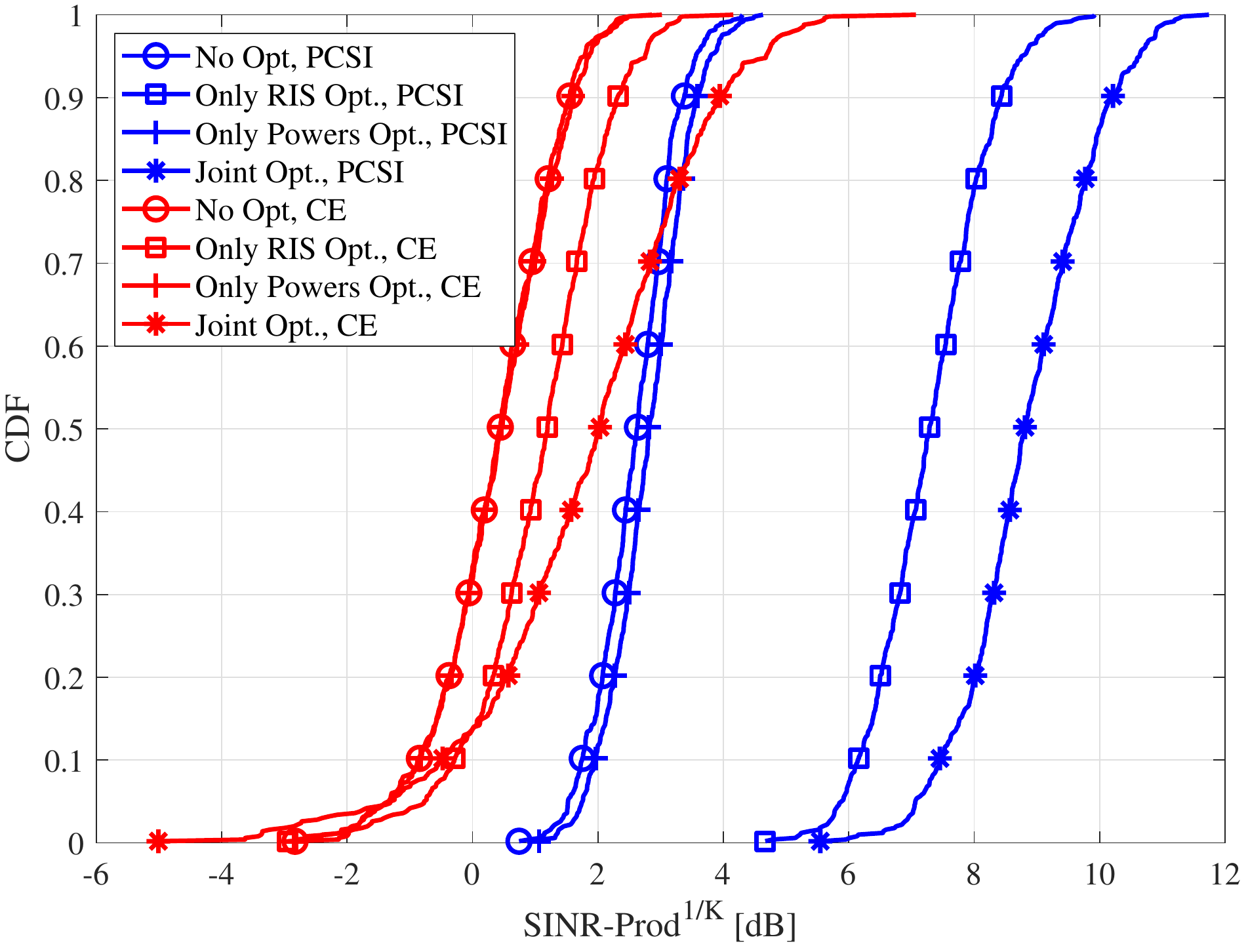}
\caption{CDFs of the geometric mean of the SINRs in the multiuser scenario.}
\label{Fig:SNR_MultiUser_Opt}
\end{figure}

Fig. \ref{Fig:SNR_MultiUser_Opt} reports the CDFs of the geometric mean of the SINRs in the multiuser scenario obtained with the joint passive beamforming design and power allocation (Joint Opt.) technique reported in Section \ref{Joint_Resource}. We compare these results with the ones obtained when only the RIS phase shifts are optimized (Only RIS Opt.), i.e. with a uniform power allocation,  and with the ones obtained when only the transmit powers (Only Powers Opt.) are optimized, assuming a random configuration of the phase shifts.

Both figures show that the proposed resource allocation procedure bring noticeable performance improvements with respect to the case in which no optimization procedure is carried out. It is also seen that there is a noticeable gap due to the lack of perfect CSI, especially in the multiuser case, but this is also due to the fact that we are considering a highly loaded system, i.e. we have $K=10$ users served with $N_B=16$ antennas at the BS. Other results, not shown here due to lack of space, have shown that performance noticeably improves if $N_B$ is increased. 

\section{Conclusion}
For a wireless network assisted by a RIS, resource allocation procedures have been proposed. In particular, the paper tackles the problem of SNR maximization with respect to  the RIS configuration and to the BS beamformer for the case in which SNR is to be maximized in a single-user setting. Moreover, for a multiuser scenario, the geometric mean of the SINRs is maximized with respect to the transmit power vector and to the RIS configuration. The obtained results are encouraging, even though there is a noticeable gap between the performance obtained in the case of perfect CSI and that obtained in the case of imperfect channel knowledge. Further research is thus needed in order to come up with high performance channel estimation algorithms.

\section*{Acknowledgment}
This paper was supported by HiSilicon Technologies  through cooperation agreement
YBN2018115022.

\bibliographystyle{IEEEtran}
\bibliography{LISreferences}

\end{document}